\documentstyle[12pt]{article}
\begin{document}
\begin{flushright}
Preprint IC/98/93 \\
 {\bf hep-th/9807202}\\ 
July 28, 1998
\end{flushright}

\begin{center}
{\Large \bf 
Superembedding approach and generalized action in String/M-theory}
\footnote{
Invited talk at the International 
Seminar "Supersymmetries and Quantum Symmetries" (June 22-26, 1997) 
dedicated to the memory of Professor Victor I. Ogievetsky. To be published in 
the Proceedings of the Seminar}. 

\bigskip
 
\bigskip

{\large \bf Igor\,Bandos}

\bigskip

{\sl Institute for Theoretical Physics, \\ 
NSC Kharkov Institute of Physics and Technology,\\ 
310108 Kharkiv, Ukraine \\ 
and \\
the Abdus Salam ICTP, \\ 
Strada Costiera 11, \\ 
34100 Trieste, Italy}

\end{center}

\bigskip

\bigskip


\begin{abstract}
A brief introduction to superembedding approach (SEA) in its variant 
based on the 
generalized action principle (GAP) 
for super-p-branes is given. 
A role of harmonic variables for Lorentz 
group is stressed. 
A relation of the GAP with complete superfield actions 
is noted. 
 Recent applications in studying of  
Dirichlet branes (super--Dp--branes)  
and M-branes are discussed. 
\end{abstract}

\bigskip

\bigskip 

1996 was the hard year for our Science. 
In January of this year we had lost our teacher Dmitrij V. Volkov 
and, in March, Victor I. Ogievetsky left us. 

In this contribution I present a brief description of 
generalized action principle (GAP) and superembedding approach (SEA) 
for supersymmetric extended objects. 
We have  proposed them and elaborated for $D=10$ superstrings 
(called now fundamental strings) and $D=11$ supermembrane 
(called now M2-brane) in collaboration with D.V. Volkov 
\cite{bpstv,bsv,Volkov}.  
On the other side, they are  based on the works of D.V. Volkov 
\cite{stvz} on doubly supersymmetric twistor-like approach, 
unify the latter  with 
the Lorentz harmonic approach \cite{bzp} and, thus,  
uses essentially the concept of harmonic variables, 
developed by V.I. Ogievetsky with collaborators \cite{gikos}.
So the subject of my talk originates from the work of both these 
great scientists. 

Recently the superembedding approach has been applied for investigation of 
 super-D-branes \cite{hs96,bst,abkz}, super-M5-brane 
\cite{hs5,hsw} 
as well as 
intersecting branes \cite{chsw} and  brane models of gauge theories 
\cite{w}. A derivation of  brane action from superembedding equations, 
which can be regarded as an inversion of the line of the 
GAP approach, has been proposed in \cite{hrs98}.
The GAP for $D=11$ supermembrane \cite{bsv} 
in $AdS_4 \times M_7$ background has been used to obtain  
the supersymmetric $Osp(8|4)$ singleton action \cite{AFFFT}.

In this talk I  
review the main ingredients of the SEA  
and the GAP using relatively simple example 
of  $D=10$ heterotic string  
and 
describe briefly  achievements and  problems   
of the SEA  in studying 
String/M-theory.

\section{Generalized action for $D=10$ heterotic string} 

{\bf 1.1.} One of the main ingredients of the generalized action (GAP) is the 
{\bf Lagrangian form }
defined on the {\bf world volume superspace} 
$$\Sigma^{(p+1|n)} = \{ (\zeta^M )\} = \{ (\xi^m , \eta^{q} )\}$$ of the super-p-brane. 
For  $D=10$ heterotic string ($\underline{m}=0,\ldots, 9;~p=1; ~m=0,1;~q=1,\ldots 8$) 
the Lagrangian two-- form 
\footnote{For simplicity we restrict ourself by the case of flat 
target superspace.  
The generalization for supergravity background is straightforward 
\cite{bst}.}
\begin{equation}\label{4}
{\cal L}_2 = 
{1 \over 2} E^{++} \wedge E^{--} 
-i \Pi^{\underline{m}} \wedge d\Theta \Gamma_{\underline{m}} \Theta
+ {1 \over 2} E^{--}
\wedge \Psi^I_{-} d \Psi^I_{-}, \qquad 
\end{equation} 
$$
I=1,\ldots , 32
$$
is constructed from  the pull-backs of the basic one forms of 
target superspace 
$$\Pi^{\underline{m}}= dX^{\underline{m}} - i d\Theta
\Gamma^{\underline{m}} \Theta, \qquad    d\Theta^{\underline{\mu}}
$$
onto the world sheet superspace 
$\Sigma^{(1+1|n)} = \{ (\zeta^M )\} = \{ (\xi^{(\pm\pm)} , \eta^{q} )\}$
$$\Pi^{\underline{m}}= d\zeta^M \Pi_M^{\underline{m}} \equiv 
d\zeta^M (\partial_M X^{\underline{m}}(\xi^{(\pm\pm )}, \eta^q) 
- i\partial_M \Theta \Gamma^{\underline{m}} \Theta ) , $$ 
$$d\Theta^{\underline{\mu}} = d\zeta^M \partial_M \Theta^{\underline{\mu}} 
(\xi^{(\pm\pm )}, \eta^q)~$$ 
 and  heterotic fermion  1--form 
$$
\Psi^I_{-} d \Psi^I_{-} = - d\zeta^M  (\partial_M \Psi^I_{-})  
\Psi^I_{-}(\xi^{(\pm\pm )}, \eta^q)
$$
{\bf by the use of the external product of the superforms only}.

Supervielbeine  of flat target superspace
$E^{\underline{A}} =(E^{\underline{a}}, E^{\underline{\alpha }})$ 
\begin{eqnarray}\label{9}
E^{\underline{a}} = (E^{\pm\pm}, E^i)
= \Pi^{\underline{m}} u_{\underline{m}}^{\underline{a}} 
= ( \Pi^{\underline{m}} u_{\underline{m}}^{\pm\pm},
\Pi^{\underline{m}} u_{\underline{m}}^{i})
\\ 
E^{\underline{\alpha}} = (E^{+ q}, E^{- \dot{q}})
= d\Theta^{\underline{\mu}} v_{\underline{\mu}}^{\underline{\alpha}} =
(d\Theta^{\underline{\mu}} v_{\underline{\mu}q}^{~+},
d\Theta^{\underline{\mu}} v_{\underline{\mu}\dot{q}}^{~-}) \nonumber 
\end{eqnarray}
distinct from the standard one ($\Pi^{\underline{m}}, 
d\Theta^{\underline{\mu}})$
 by a Lorentz rotation.
The vector and spinor representations of this $SO(1,D-1)$ transformation
are given by the matrices
$$
u^{~\underline{a}}_{\underline{m}} =
(u^{\pm\pm}_{\underline{m}}, u^{~i}_{\underline{m}}) ~ \in ~ SO(1,9)
\qquad 
v^{~\underline{\alpha}}_{\underline{\mu}} =
(v^{~+}_{\underline{\mu}q}, v^{~-}_{\underline{\mu}\dot{q}})
~ \in ~ Spin(1,9)
$$
\begin{equation}\label{11}
\Leftrightarrow ~~~~
u^{~\underline{a}}_{\underline{m}} \eta^{\underline{m}\underline{n}}
u^{~\underline{b}}_{\underline{n}} = \eta^{\underline{a}\underline{b}}
~~~~\Leftrightarrow ~~~~
 \cases {  
 u^{++}_{\underline{m}} u^{++\underline{m}} = 0, ~~ 
 u^{--}_{\underline{m}} u^{--\underline{m}} = 0, \cr
 u^{--}_{\underline{m}} u^{++\underline{m}} = 2, \cr 
 u^{i}_{\underline{m}} u^{j\underline{m}} = - \delta^{ij}, ~~ 
 u^{\pm\pm}_{\underline{m}} u^{i\underline{m}} = 0, \cr }
\end{equation}
 (vector and spinor  ${{SO(1,9)} \over {SO(1,1)\times SO(8)}}$  
 {\bf Lorentz harmonics}, see \cite{bpstv} and refs. 
 therein). They  are related by 
$$u^{~\underline{a}}_{\underline{m}}
\Gamma^{\underline{m}}_{\underline{\mu}\underline{\nu}}
= v_{\underline{\mu}}^{~\underline{\alpha}}
\Gamma^{\underline{a}}_{\underline{\alpha }\underline{\beta }} 
v^{~\underline{\beta}}_{\underline{\nu}},  
\qquad u^{~\underline{a}}_{\underline{m}}
\Gamma_{\underline{a}}^{\underline{\alpha }\underline{\beta }}
= v_{\underline{\mu}}^{~\underline{\alpha}}
\Gamma_{\underline{m}}^{\underline{\mu }\underline{\nu }}
v^{~\underline{\beta}} _{\underline{\nu}}, 
$$
\begin{equation}\label{8}
\Leftrightarrow \qquad \delta_{qp}
u^{~++}_{\underline{ m}} = v^+_q \Gamma_{\underline{
m}} v^+_p , \qquad 
\delta_{\dot{q} \dot{p}} u^{--}_{\underline{ m}} =
v^{-}_{\dot q}
\Gamma_{\underline{m}}
v^{-}_{\dot p} , \qquad 
\gamma ^{i}_{q \dot p}
u^{i}_{\underline{ m}}
= v^+_q
\Gamma_{\underline{ m}}
v^{-}_{\dot p}    , 
\end{equation}
(where $\Gamma_{\underline{m}}$ and
$\gamma ^{i}_{q \dot p}$
 are the $SO(1,9)$ and
 $SO(8)$  $\gamma$--matrices). 
Their differentials 
 \begin{equation}\label{81}
 d u^{~\underline{a}}_{\underline{m}} =
 u^{~\underline{b}}_{\underline{m}}
 \Omega^{~\underline{a}}_{\underline{b}} (d)
 \qquad  \Leftrightarrow \qquad
 \cases {
 du^{\pm\pm}_{\underline{m}}
 = \pm u^{\pm\pm}_{\underline{m}}  \Omega^{(0)} (d)
 + u^{~i}_{\underline{m}} \Omega^{\pm\pm i} (d ) , 
 \cr
 d u^{i}_{\underline{m}} = - u^{j}_{\underline{m}}  \Omega^{ji} +
 {1\over 2} u_{\underline{m}}^{\pm\pm} \Omega^{\mp\mp i} (d)
  \cr }
\end{equation}
 \begin{equation}\label{82}
 d v^{~\underline{\alpha}}_{\underline{\mu}} = 
 1/4 v^{~\underline{\beta}}_{\underline{\mu}}
 (\Gamma_{\underline{a}\underline{b}})^{~\underline{\alpha}}_{\underline{\beta}}
 \Omega^{~\underline{a}\underline{b}} (d)
  \qquad
 \end{equation} 
are expressed in terms of the
$so(1,D-1)$ valued Cartan $1$--forms
\begin{equation}\label{Cfv}
\Omega^{\underline{a}\underline{b}} = -
\Omega^{\underline{b}\underline{a}} =
\left(
\matrix{ \Omega^{ab} & \Omega^{aj} \cr
       - \Omega^{bi} & \Omega^{ij} \cr}
        \right)
= u^{~\underline{a}}_{\underline{m}} d u^{\underline{b}\underline{m}}
= {1 \over 16} 
dv^{~\underline{\alpha}}_{\underline{\mu}}
(\Gamma^{~\underline{a}\underline{b}})_{\underline{\alpha }}^{\underline{\beta}}
v^{~\underline{\mu }}_{\underline{\beta}}
\end{equation}

\bigskip

 Integrating the Lagrangian form (\ref{4}) over 
{\bf pure bosonic} world sheet
\begin{equation}\label{ac0}
S_0 =
\int_{{\cal M}^{1+1}_0}  
{\cal L}_{2} \equiv \int d^2\xi ~~\epsilon^{nm} ~~({\cal L}_{2})_{mn} 
\vert_{\eta^q =0}
\end{equation}
gives the usual (component) superstring action in the first order form 
\cite{bzp,bpstv}).
The equations of motion following from the functional 
(\ref{ac0})\footnote{
These nontrivial equations are produced by 
variations of 
 Lorentz harmonic variables 
$\delta S /i_\delta \Omega^{\pm\pm i} \equiv u^{\underline{m}i}\delta S / 
\delta u^{\underline{m}\mp\mp} = 0~~$ (\ref{Ei}),  heterotic fermions 
$\delta S / 
\delta \Psi^{I}_- = 0~~$ 
(\ref{Psieq}),  
and superspace coordinate fields 
$\delta S /i_\delta E^i \equiv u^{\underline{m}i}\delta S / 
\delta X^{\underline{m}} = 0~~$ (\ref{E-q}), 
$~~\delta S /i_\delta E^-_{\dot{q}} \equiv 
v^{+\underline{\mu}}_{\dot{q}} \delta S / 
\delta \Theta^{\underline{\mu}} = 0$ (\ref{Beq}) respectively.}
\begin{equation}\label{Ei}
E^i \equiv \Pi^{\underline{m}} u^i_{\underline{m}} = 0. 
\end{equation}
\begin{equation}\label{Psieq}
E^{--} \wedge {\cal D}\Psi_-^I =0 \qquad \Rightarrow 
\qquad {\cal D}\Psi_-^I = E^{--} 
{\cal D}_{--}\Psi_-^I 
\end{equation}   
\begin{equation}\label{E-q}
E^{-}_{\dot{q}}=( E^{++} - {1 \over 2} \Psi_-^Id\Psi_-^I ) 
\psi^{~~-}_{++\dot{q}} 
= (E^{++} - E^{--} {1 \over 2} \Psi_-^I{\cal D}_{--}\Psi_-^I)
\psi^{~~-}_{++\dot{q}}
\end{equation}
\begin{equation}\label{Beq}
E^{--} \wedge \Omega^{++i} - ( E^{++} - \Psi_-^Id\Psi_-^I ) 
\wedge \Omega^{++i} + 4i E^{+}_q \wedge E^{-}_{\dot{q}} \gamma^{i}_{q\dot{q}} = 0
\end{equation} 
can be reduced to the standard 
superstring equations upon eliminating the auxiliary fields (harmonics).
A part of variations 
does not produce independent equations. They, hence, can be identified 
with the parameters of the gauge symmetries:  
$$ i_\delta \Omega^{(0)} = {1\over 2} u^{--}{}^. \delta u^{++} \qquad
\rightarrow \qquad SO(1,1) ,$$ 
$$ i_\delta \Omega^{ij} = u^i{}^. \delta u^j \qquad \rightarrow \qquad SO(8) , $$ 
$$ i_\delta E^a \rightarrow \qquad 'b-symmetry'~ or~ 
reparametrization,$$ 
$$\kappa^+_q=i_\delta E^+_q \qquad \rightarrow \qquad \kappa - symmetry.$$ 

\bigskip

 The {\bf generalized action} 
for $D= 10$ heterotic string
\begin{equation}\label{1}
S =
\int_{{\cal M}^{1+1}}  {\cal L}_{2} \equiv 
\int_{{\cal M}_0^{1+1}} {\cal L}_{2}\vert_{\eta^q = \eta^q (\xi )}
\end{equation}
\cite{bsv}, see \cite{rheo} for supergravity)
is given by the integral of the Lagrangian 2--form (\ref{4})
{\bf over
 arbitrary $2$ -- dimensional bosonic surface}
$${\cal M}^{1+1} = \{ (\xi^{(\pm\pm)} , \eta^{(+)q} ):
\eta^{(+)q}= \eta^{(+)q}(\xi)\} \qquad \in \qquad \Sigma^{(1+1|n)}
$$
{\bf in the world volume superspace}  
$\Sigma^{(1+1|n)}= \{ (\xi^{(\pm\pm)} , \eta^{(+)q} )\} $. 
Hence, in  the functional (\ref{1}) all the variables 
shall be considered as
world volume superfields
but, taken on the surface ${\cal M}^{1+1}$ (i.e. with 
$\eta^q = \eta^q(\xi )$):  
$$
X^{\underline{m}} = X^{\underline{m}} (\xi , \eta (\xi)),~~~ 
\Theta^{\underline{\mu}} = \Theta^{\underline{\mu}} (\xi , \eta (\xi)) ,~~~ 
u^a_{\underline{m}}= u^a_{\underline{m}}(\xi , \eta (\xi)),~~~ 
\Psi_-^I= \Psi_-^I (\xi , \eta (\xi)) .
$$

The variation of the GAP (\ref{1})
should vanishes for arbitrary variations of the (super)fields involved
{\sl as well as for arbitrary variations of the surface} 
${\cal M}^{1+1}$ (i.e.  $\delta S /\delta \eta (\xi) = 0 $). 
For the 
Lagrangian form under consideration it can be proved 
that the latter 
variation 
does not lead to new
equations of  motion. 
Instead, the arbitrariness of the surface ${\cal M}^{(p+1)}$ 
provides the possibility
to regard the 'field'  equations  
(\ref{Ei}), (\ref{Psieq}), (\ref{E-q}), (\ref{Beq}), 
{\bf as equations for forms and superfields defined on the whole world volume 
superspace} 
$\Sigma^{(1+1|8)}$.

\section{Superembedding equations of heterotic string and complete superfield action}

Thus the GAP produces {\bf formally the 
same equations} (\ref{Ei}), (\ref{Psieq}), (\ref{E-q}), (\ref{Beq}). 
However, now they 
{\bf can be 
considered 
as equations for superfields and differential forms on the world volume 
superspace} 
$\Sigma^{(1+1|8)}$.

A pragmatic way to find that a Lagrangian form ${\cal L}$ is proper 
for constructing 
the GAP is to prove that corresponding 
equations 
are selfconsistent as equations on the  world volume 
superspace $\Sigma^{(p+1|n)}$ of the super-p-brane, i.e. form 
{\sl the free differential algebra on this superspace}    
 \cite{rheo}). 

The Eqs. (\ref{Ei}), (\ref{Psieq}), (\ref{E-q}), 
(\ref{Beq}) are selfconsistent on $\Sigma^{(1+1|8)}$
and describe the  
{\bf minimal (on-shell) embedding}  
of the heterotic string world volume superspace 
$\Sigma^{(1+1|8)}$ into the flat  $D=10, N=1$  superspace. 
Not all of them are independent. 
Eqs.  (\ref{E-q}), and the bosonic component of 
(\ref{Psieq}) $D_{+q}\Psi^I_-=0$ are {\bf independent  dynamical 
superfield equations} while Eq. (\ref{Beq}) and Grassmann component 
of (\ref{Psieq}) $D_{++}\Psi^I_-=0$  appear as their consequences.

The {\bf off--shell superembedding} $\Sigma^{(1+1|8)}$ is specified by 
the equation (\ref{Ei}) only. 
The Eq. (\ref{Ei}) can be recognized as the {\sl geometrodynamic equation} 
\cite{stvz,dghs92,bpstv} and refs.therein) or superembedding condition 
\cite{hs96,hs5}). 
To justify the equivalence of (\ref{Ei}) with the standard 
form of superembedding 
equations 
\begin{equation}\label{EiPi}
E^i \equiv \Pi^{\underline{m}} u^i_{\underline{m}} = 0 
\qquad \Leftrightarrow  \qquad \Pi_{+q}^{\underline{m}} \equiv 
{\cal D}_{+q}X^{\underline{m}} - i {\cal D}_{+q}\Theta \Gamma^{\underline{m}}
\Theta, 
\end{equation}
we have to take into account 
the freedom in choice of
 the intrinsic supervielbein of the world volume superspace 
$$
e^A = (e^{\pm\pm}, e^{+q}) = d\xi^{(\pm\pm)} e_{(\pm\pm)}^A + d\eta^{(+)q} 
e_{(+)q}^A ,
$$ 
$$(d=e^A{\cal D}_A =  e^{+q}{\cal D}_{+q}+ e^{\pm\pm}{\cal D}_{\pm\pm}), $$ 
as well as a freedom to adapt the moving frame (Lorentz harmonics) to the 
world volume (see \cite{bpstv,hrs98})
\footnote{In the presence of heterotic fermions the most convenient choice 
of the world volume supervielbein is 
$$ 
e^{--} = E^{--} \equiv \Pi^{\underline{m}} u^{--}_{\underline{m}},~~~ e^{++} = E^{++}- a\Psi_-^Id\Psi_-^I, 
~~~ e^{+q}= E^{+q} \equiv d\Theta^{\underline{\mu}} 
v^{~+}_{\underline{\mu}q}
$$ 
with $a=1/2$. The harmonic frame can be chosen in such a way that  
$\Pi_a^{\underline{m}} u_{\underline{m}}^{i} = 0$ holds.}.   

\bigskip

As it is known \cite{dghs92}, 
{\bf for heterotic string  the geometrodynamic condition (\ref{EiPi})    
does not contain equations of motion 
among its consequences}.
Indeed, from the integrability condition for Eq. (\ref{EiPi})  one obtains 
\begin{equation}\label{E-qoff}
E^-_{\dot{q}} = E^{++} \psi^{~~-}_{++\dot{q}} + E^{--} \psi^{~~-}_{--\dot{q}}
\end{equation}
while the superfield equations of motion (\ref{E-q}) specifies 
$$\psi^{~~-}_{--\dot{q}} = -{1 \over 2} \Psi_-^I{\cal D}_{--}\Psi_-^I\psi^{~~-}_{++\dot{q}}.$$

\bigskip

{\bf In the absence of heterotic fermions}  
the external derivative of the Lagrangian form
(\ref{4}) 
\begin{equation}\label{dL2}
d{\cal L}_2 \vert_{\Psi_-^I=0}= 
-2 i E^-_{\dot{q}} \wedge E^-_{\dot{q}} \wedge E^{++} 
+{1 \over 2} E^i \wedge ( E^{\mp\mp} \wedge \Omega^{\pm\pm i} 
- 4i E^+_q \wedge E^-_{\dot{q}}\gamma^i_{q\dot{q}})
\end{equation}
evidently vanishes as a consequence of the geometrodynamic condition 
(\ref{EiPi}) only. As the latter has no dynamical content, this statement 
means the {\sl off-shall superdiffeomorphism invariance} of the 
GAP (in the {\sl rheonomic sense}). 

This provides the possibility to construct the 
{\bf complete superfield action} 
using the 
GAP Lagrangian 
form 
${\cal L}_2 \equiv 1/2 d\zeta^M \wedge d\zeta^N {\cal L}_{NM}$
\begin{equation}\label{ssac}
S_{superfield} = \int d^{2}\xi d^8 \eta ~~
(P^{+q}_{\underline{m}} \Pi^{\underline{m}}_{+q} + 
~P^{MN} ~({\cal L}_2 - dY)_{NM}).
\end{equation}
Here $P^{+q}_{\underline{m}}$, $P^{MN}$ are Lagrangian multipliers and
$Y=d\zeta^M Y_M$ is an auxiliary 1--form superfield. 
The functional (\ref{ssac}) is just the superfield action for heterotic string 
discovered in \cite{dghs92}, where the GAP Lagrangian form 
${\cal L}_2$ was called 'Wess--Zumino 2--form'. 
It is worth mentioned that in all the superfield functionals for 
superbranes (see \cite{stvac} and refs. in \cite{bpstv}) the so-called 
{\bf 'Wess-Zumino form' is} nothing else 
then {\bf the GAP Lagrangian form} for corresponding brane.

\bigskip

When {\bf the heterotic fermions are present}, 
the derivative of the Lagrangian form is
$$
d{\cal L}_2 = d{\cal L}_2 \vert_{\Psi_-^I=0} +  
 i E^-_{\dot{q}} \wedge E^-_{\dot{q}} \wedge \Psi_-^Id\Psi_-^I
+{1 \over 2} E^{--} \wedge {\cal D}  \Psi_-^I\wedge {\cal D} \Psi_-^I
$$   
To write the complete superfield action, the inputs from heterotic fermions 
has 
to be separated and considered with some care \cite{st,is,h94,bcsv}. 

Thus  the GAP produces superembedding equations 
together with their consequences (including proper equations of motion). 
This can simplify essentially the studying of brane superembeddings.  
On the other hand, when 
the Lagrangian form is closed on the surface of nondynamical equations, 
GAP can be used for construction of the complete superfield action.

\section{Generalized action and superembedding approach to D-branes and 
M-branes}

The superembedding approach (SEA) demonstrated its strength in studying 
the  new 
objects of String/M-theory: $D=10$ Dirichlet superbranes (super-Dp-branes) 
and M5-brane. The equations of motion for these objects had been 
obtained in the frame of superembedding approach \cite{hs96,hs5} before the 
covariant action functionals were constructed 
\cite{Dbrac,aps,bt,blnpst,schw5}. 

The complete bridge between the covariant actions and the SEA description 
can be build by constructing the GAP. 

The GAP Lagrangian form  
in all the cases includes {\sl the same Wess--Zumino term} 
${\cal L}_{p+1}^{WZ}$ as 
the 'standard' (component) 
action does. 
So the only problem is to write down the 'kinetic' part 
(in terms of differential forms,  
without any use of the Hodge operation). 
This goal is achieved {\bf using Lorentz harmonics} 
$u^{\underline{a}}_{\underline{m}} =(u^{{a}}_{\underline{m}}, 
u^{{i}}_{\underline{m}})~~\in~~SO(1,D-1)$ ($a=0,\ldots p,~~i=1,\ldots (D-p-1)$)
providing the possibility to adapt  
the bosonic component of target superspace supervielbein $E^{\underline{a}}$ 
to the superembedding of the world volume superspace 

$$
E^{\underline{a}}= (E^a, E^i), \qquad 
E^a \equiv \Pi^{\underline{m}} u^a_{\underline{m}}, \qquad  
E^i \equiv \Pi^{\underline{m}} u^i_{\underline{m}} 
$$

The GAP for {\bf M2-brane (supermembrane)} 
is known since 1995 
\cite{bsv}  
and 
is based on the Lagrangian form
\begin{equation}\label{LM2}
{\cal L}_{3}^{M2} = E^{\wedge 3} + {\cal L}_3^{WZ~M2}, \qquad 
with \qquad E^{\wedge 3} \equiv {1 \over 3!} \epsilon_{abc} 
E^a \wedge E^b \wedge E^c    
 \end{equation}
It produces the superembedding equation $$E^i=0,$$ its consequence 
$$E^{\alpha}_{\dot{q}} \equiv d\Theta^{\underline{\mu}} 
v^{~\alpha}_{\underline{\mu}\dot{q}} = E^a \psi^{~\alpha}_{a\dot{q}}$$ 
as well as the superfield equations of motion 
$$\psi^{~\alpha}_{a\dot{q}} \gamma^a_{\alpha\beta} = 0 . $$ 
 
\bigskip

The Lagrangian form for $D=10$ {\bf super-Dp-branes}  
\begin{equation}\label{LDp}
{\cal L}^{Dp}_{p+1} = E^{\wedge (p+1)} {\sqrt{-det(\eta_{ab}+F_{ab})}} +
Q_{p-1} \wedge (dA-B_2 - {1\over 2} E^b \wedge E^c F_{cb}) 
+ {\cal L}_{p+1}^{WZ~Dp}  
\end{equation} 
\cite{bst}  includes  the auxiliary tensor field $F_{ab}=-F_{ba}$ as well as 
$(p-1)$--form Lagrange multiplier $Q_{p-1}$. It produces 
the generalized {\bf gauge field constraints}  
$$ ~dA-B_2 = {1\over 2} E^b \wedge E^c F_{cb},$$ 
which complete the 
geometrodynamic
condition $$E^i=0$$ up to the complete set of superfield equations for 
{\sl any} 
$Dp-brane$ (cf. with \cite{hs96,w}). 
The geometrodynamic equation $E^i=0$ 
as well as its consequence ({\sl fermionic superembedding condition}) 
$$E^{\alpha 2}_q = E^{\beta 1}_q h_{\beta}^{~\alpha} + E^a \psi_{a q}^{\alpha}$$
and equations of motion 
$$\psi_{aq}^{~\alpha} (\eta + F)^{-1~ab} (\gamma_b )_{\alpha \beta}=0$$ follow from the GAP (\ref{LDp}) as well. 

\bigskip

It is worth mentioned that 
the {\bf spin--tensor  field}  $h_{\beta}^{~\alpha}$ involved into 
the fermionic superembedding conditions 
$E^{\alpha 2}_q = E^{\beta 1}_q h_{\beta}^{~\alpha} + E^a \psi_{a q}^{\alpha}$
\cite{hs96},(1997)) is {\bf Lorentz group valued} 
$$h_{\beta}^{~\alpha}~\in~Spin(1,p)$$
and provides the {\bf spinor representation 
for the Cayley image} $$ k_b^{~a}\equiv (\eta - F)_{bc}
(\eta + F)^{-1~ca}~\in ~ SO(1,p)$$ 
 of the gauge field strength $F_{ab}$
\cite{abkz}
$$
~h_{\beta}^{~\beta '} \gamma^a_{\beta '\alpha '} h_{\alpha }^{~\alpha '} = 
\gamma^b_{\beta \alpha } k_b^{~a}.$$ 

\bigskip

Returning to M-branes, note that the GAP for $D=11$ massless 
superparticle or {\bf M0-brane}  
\cite{bt} 
can be constructed on the basis of  
Lagrangian 1--form 
$$
{\cal L}_1 = P_{--} v^{~-}_{\underline{\mu}A} v^{~-}_{\underline{\nu}A} 
\Gamma_{\underline{m}}^{\underline{\mu}\underline{\nu}} 
\Pi^{\underline{m}}, \qquad  
 (A=1,...,16) $$ 
\cite{BL} 
where (cf.  \cite{bn} )
$P_{--}$ is Lagrange multiplier, and 
$$v^{~\underline{\alpha}}_{\underline{\mu}}= (v^{~-}_{\underline{\mu}A},  
v^{~+}_{\underline{\mu}A}) ~\in ~Spin(1,10)$$ are 
spinor Lorentz harmonics parametrizing the coset 
$${SO(1,10) \over SO(1,1) \otimes SO(9) \subset\!\!\!\!\!\!\times K_9}$$ 

\bigskip

One of the {\bf open problems} is 
the construction of the GAP for M5-brane. 
The superembedding equations for the M5-brane are known 
\cite{hs5}, 
produces the same field equations as the covariant action 
\cite{blnpst,schw5} 
 and found many physical applications \cite{w}. 

However, the natural candidate for the GAP Lagrangian form 
${\cal L}_6^{M5}$ (see \cite{bpst}), whose integration over the 
bosonic world volume provides the first order form for the action
\cite{blnpst},  
produces 
(in addition to the 
superembedding equations $$E^i=0, \qquad  
E_{\alpha q}= E^{\beta}_q h_{\beta \alpha} + E^a \psi_{a\alpha q}$$  
\cite{hs5})
the equation $$da=E^a u_a ,$$  
whose superspace 
extension 
has only trivial solutions.  
As the presence of the closed form $da$ is characteristic 
for the PST approach, developed for covariant  Lagrangian description of 
self--dual (chiral) fields 
\cite{pst} and used in 
\cite{blnpst},
 the problem can be formulated in more general way as looking for  
a consistent  unification of the GAP with the PST approach. 
Another  possible way consists in searching for reformulation of 
the M5--brane action  in terms of the spin--tensor field 
$$h_{\alpha\beta}\equiv h^{abc} 
\tilde{\gamma}_{abc}^{\alpha\beta}$$ (involved into the fermionic 
superembedding condition 
$E_{\alpha q}= E^{\beta}_q h_{\beta \alpha} + E^a \psi_{a\alpha q}$) instead of $H_3=dB_2 -A_3$ and the PST scalar $a$. 
Similar reformulation of D3-brane action was found recently \cite{BK}.

Another relevant  problem for further study consists in   
searching for SEA and GAP description for $D=11$ KK monopole 
\cite{KK} and M--brane \cite{M9}.

{\bf Acknowledgements.} Recent results reviewed in this talk were obtained in 
fruitful collaboration with  V. Akulov, 
W. Kummer, K. Lechner, A. Nurmagambetov, P.Pasti, 
D. Sorokin, M. Tonin, V. Zima. Author is thankful to 
D. Sorokin, M.Tonin and  
E. Sezgin for useful 
discussions. 
Author is grateful to Prof. M. Virasoro and  
Prof. R. Randjbar-Daemi for hospitality at the ICTP were this work was 
completed.

\end{document}